\documentclass[preprint,prd,superscriptaddress,showpacs,aps]{revtex4}

\usepackage{amsmath}
\usepackage{amsbsy}
\usepackage{amsthm}
\usepackage{amssymb}
\usepackage{latexsym}
\usepackage[dvips]{graphicx}
\usepackage{epsfig,enumerate,color,rotating,colordvi,psfrag,epsf}

\begin{document}

\input{epsf}

\title{Hadron spectra from a non-relativistic model\\
with confining harmonic potential\\ \vspace{0.4cm} \small{Espectros 
de Hadrons a partir de um modelo n\~ao relativ\'{\i}stico \\ com 
potencial harm\^onico confinante}} 

\author{Eduardo Cuervo-Reyes}
\affiliation{Instituto Superior de Ciencias y Tecnolog\'{\i}a 
Nucleares-ISCTN,\\
Ave. Salvador Allende y Luaces, Quinta de los Molinos, La Habana, Cuba}

\author{Marcos Rigol}
\affiliation{Institut f\"ur Theoretische Physic III, Universit\"at 
Stuttgart,\\ Pfaffenwaldring 57, D-70550 Stuttgart, Germany}
\affiliation{Centro Brasileiro de Pesquisas F\'{\i}sicas-CBPF,
Rua Dr. Xavier Sigaud 150, 22229-180, Rio de Janeiro, Brazil}

\author{Jesus Rubayo-Soneira}
\affiliation{Instituto Superior de Ciencias y Tecnolog\'{\i}a 
Nucleares-ISCTN,\\
Ave. Salvador Allende y Luaces, Quinta de los Molinos, La Habana, Cuba}

\vspace{-5cm}

\begin{abstract}

Hadron spectra and other properties of quark systems are studied in the 
framework of a non-relativistic spin-independent phenomenological 
model. The chosen confining potential is harmonic, which allowed us to 
obtain analytical solutions for both meson and baryon (of equal constituent 
quarks) spectra. The introduced parameters are fixed from the low-lying 
levels of heavy quark mesons. The requirement of flavor independence is 
imposed, and it restricts the possible choices of inter-quark potentials. 
The hyper-spherical coordinates are considered for the solution of the 
three-body problem.\\

Espectros de Hadrons e outras propriedades de sistemas de quarks s\~ao estudados 
do ponto de vista de um modelo fenomenol\'ogico n\~ao-relativ\'{\i}stico 
independente de spin. O potencial confinante escolhido \'e harm\^onico, o 
qual nos permite obter solu\c{c}\~oes anal\'{\i}ticas tanto para os espectros 
dos m\'esons quanto para os dos b\'arions (de iguais quarks constituintes). Os 
par\^ametros introduzidos s\~ao fixos a partir dos primeiros estados excitados 
dos m\'esons pesados. A condi\c{c}\~ao de independ\^encia de sabor \'e imposta, 
o que restringe as poss\'{\i}veis escolhas de potenciais inter-quarks. As 
coordenadas hiper-esf\'ericas s\~ao consideradas para a solu\c{c}\~ao do 
problema de tr\^es corpos.

\end{abstract}


\maketitle

\section{Introduction}

The study of the fundamental or constituent blocks of matter has been for 
long time a fascinating field in physics. With the pass of the years new 
fundamental blocks or particles have appeared, modifying old concepts. 
For example the atom, that was initially supposed to be fundamental, was 
found to be formed by the nucleus and electrons, the nucleus was found 
to be formed by neutrons and protons (nucleons) and finally the nucleons 
to be formed by quarks. 

Nowadays it is believed that the theory describing the interactions between 
fundamental particles is the Standard Model. This model has a 
$SU\left( 3 \right) \times SU\left( 2 \right) \times U\left( 1 \right)$ 
symmetry. The $SU\left( 3 \right)$ component is what is called Quantum 
Chromodynamics (QCD) and it is the gauge field theory describing the strong 
interactions of quarks and gluons. The $SU\left( 2 \right) \times 
U\left( 1 \right)$ component is called Standard Electroweak Model describing 
interactions between quarks and leptons through the $W$ bosons and the photons.
There is an additional particle in the electroweak model, the neutral Higgs 
scalar that appears after the spontaneously symmetry breaking mechanism, whose 
existence has not been probed experimentally and that is a subject of a lot of 
present experimental research because it could prove the validity of the theory. 

We will be interested in the present paper in the low energy region of the 
QCD theory, in which quarks interact strongly to form bound states known as 
hadrons. When these bound states are formed by a quark and an antiquark 
($q\overline{q}$) they are called mesons; and when they are formed by 3-quark 
states ($qqq$) they are called baryons. The study of the hadron spectra is a 
fundamental and open field in theoretical physics. Up to the moment the more 
important approaches to this subject have been: Lattice QCD, QCD sum rules 
and potential models. Lattice QCD is the most fundamental approach and together 
with QCD sum rules have had good success. Potential models, although less 
fundamentals, have proved to be very useful even in non-relativistic 
approximations. Since the latest seventies a lot of attempts have been made 
in this field with very good results \cite{Eich1}.

An important question for the use of potential models is whether it is 
possible to consider hadrons as non-relativistic bound states of 
quarks. This question could be answered solving the Schrodinger equation 
assuming a $q\overline{q}$ potential and then checking if the obtained 
quark velocities are non-relativistic. For the charmonium system this was 
done in Ref.\cite{Eich2} and it was obtained a result of $\left\langle 
v^{2}/c^{2}\right\rangle =0.2$ in the ground state. For lighter hadrons 
the results are no so good and the validity of the non-relativistic 
approximation depends strongly from the interaction potential chosen. For 
example, the interaction potential proposed by De Rujula, Georgi and 
Glashow \cite{DeRuj} was shown \cite{Bhad} to be unsuitable for dynamical 
non-relativistic calculations of light hadrons. However, Martin showed 
latter \cite{Mart1,Mart2,Mart3} that a non-relativistic model with a 
power law potential is able to describe heavy quark mesons and the 
clearly relativistic $s\overline{s}$ states. The potential proposed by Martin 
had the form
\begin{equation}
\label{Mart}
V\left(r\right) =A+Br^{\alpha},
\end{equation}
\noindent where $A=-8.093\ GeV$, $B=6.898\ GeV$ and $\alpha =0.1$.

The study of baryon systems with this power law potential was done by 
Richard \cite{Rich} obtaining good results. At that time the baryon 
spectra in potential models, although rather elaborated \cite{Bar80}, was 
completely disconnected from the meson sector. Exceptions were some 
attempts to derive meson and baryon potential energies from a common 
framework as the instanton, string or bag model. The rule adopted by Richard 
for the $qq$ potential was
\begin{equation}
\label{Rich}
V_{qq}=\frac{1}{2}V_{q\overline{q}}.
\end{equation}

Other very well known phenomenological non-relativistic potential 
models have
been based in the Cornel potential \cite{Eich3,Eich2}
\begin{equation}
\label{Corn}
V\left( r\right) =A_{C}r-\frac{B_{C}}{r}+C_{C},
\end{equation}
\noindent where $A_{C}=0.1756\ GeV^{2}$, $B_{C}=0.52$ and 
$C_{C}=-0.8578\ GeV$;
and the logarithmic potential \cite{Quigg}
\begin{equation}
\label{log}
V\left( r\right) =A_{L}+B_{L}\ln \left( r\right), 
\end{equation}
\noindent where $A_{L}=-0.6631\ GeV$ and $B_{L}=0.733\ GeV$.

Finally, other two propositions of inter-quark potentials are due to 
Song and
Ling \cite{Song}
\begin{equation}
\label{song}
V\left( r\right) =A_{S}r^{\frac{1}{2}}+B_{S}r^{-\frac{1}{2}},
\end{equation}
\noindent where $A_{S}=0.511\ GeV^{\frac{3}{2}}$ and $B_{S}=0.923\ 
GeV^{\frac{1}{2}}$;
and Turin \cite{Turin}
\begin{equation}
\label{Turin}
V\left( r\right) =-A_{T}r^{-\frac{3}{4}}+B_{T}r^{\frac{3}{4}}+C_{T}.
\end{equation}
\noindent where $A_{T}=0.620\ GeV^{\frac{1}{4}}$, $B_{T}=0.304\ 
GeV^{\frac{7}{4}}$
and $C_{T}=-0.823\ GeV$.

Studies with phenomenological potentials, like the Cornell one, but 
considering a relativistic kinetic energy term, are able to describe the 
observed spectra of heavy and light hadron systems \cite{Gara,Fulch1}. 
Non-trivial connections between these relativistic potential models and 
rigorous numerical results from lattice QCD have been demonstrated 
\cite{Dunc}. Some recent works have tried to understand why the 
non-relativistic treatment works and allows useful predictions even for 
relativistic systems \cite{Lucha,Jaczko}.

In the previously mentioned potential models, central conditions have 
been the flavor independence of the potential chosen and the existence 
of a confining term. Quark masses appearing in these phenomenological 
models are the so called constituent quark masses that should not be 
confused with the current quark masses, that are the mass parameters 
appearing in the QCD Lagrangian. Constituent quark masses are bigger 
than current quark masses and it is suppose that this is due to gluonic 
condensate effects. In general constituent quark masses are considered 
as free parameters to fit in potential models, that is why different 
values are found all over the literature. Up to our knowledge there are 
two works in which constituent quark masses are calculated from QCD: one 
is due to Elias et al. \cite{Elias1,Elias2} that used an operator-product 
expansion (OPE) of approximate non-perturbative vacuum expectation 
values in the fixed-point gauge; the other work is due to Cabo and Rigol 
\cite{Cabo} in which a Modified Perturbative QCD expansion incorporating 
gluon condensation was employed \cite{Rigol}. Because there are no 
free quarks, a lot of care should be taken with the meaning of these 
quark masses.

In the present work we study the hadron spectra within a non-relativistic 
spin-independent phenomenological model, with a harmonic confining 
potential. The idea is to show with an educational perspective what can 
be done with the Schrodinger equation, the very useful for physicists 
harmonic potential, and the same number of parameters used in almost all 
the calculations, in order to understand the hadron spectroscopy. The 
harmonic potential has the great advantage that allowed us to obtain 
analytical solutions for both the meson and baryon (with equal constituent 
quarks) spectra. Although the obtained results are not as good as the 
ones obtained for the former potentials, we think that they give very 
good estimates of the hadron properties without the need of numerical 
calculations implemented for all the previously mentioned potentials. As 
we will work with a spin independent model, the magnitudes we will deal 
with will be spin averaged. For fitting our parameters spin averages of 
experimental values were calculated, but for certain resonances the 
experimental values were not at hand \cite{EuropR} and theoretical results 
obtained by Fulcher \cite{Fulch2} were used. It should be mentioned 
that in a work by Hirata et al. \cite{Hirata} the harmonic oscillator was 
employed as an unperturbed confinement potential, in the asymptotically 
free colored-quark-gluon model in which the one-gluon-exchange force 
was treated perturbatively. Later Ram and Hasala \cite{Ram} used the 
pure harmonic oscillator potential in the Klein-Gordon equation to 
determine the meson masses.

The exposition will be organized as follows. In Section \ref{2} 
meson spectra is calculated for two different potentials with harmonic 
confining terms, that allowed to obtain analytical solutions. The 
first potential is the harmonic oscillator and the second one is the 
harmonic oscillator plus a term proportional to $\frac{1}{r^{2}}$. The 
results are analyzed for the flavor dependent and independent cases. In 
Section \ref{3} we study the properties of the radial wave function, and 
some related physical quantities, at the origin. Section \ref{4} is 
devoted to the study of the baryon spectra and the summary can be found in 
Section 5. Finally, an Appendix was introduced for mathematical details 
of the three-body problem solution.

\section{Meson Spectra \label{2}}

In the present section, the non-relativistic meson spectra are 
calculated for two different potentials. The first one is the pure harmonic 
potential and in the second one a term proportional to $\frac{1}{r^{2}}$ 
is added to the harmonic oscillator, allowing to improve the short-range 
interaction. That is the potentials considered are:
\begin{equation}
\label{V1}
V_{1}\left( r\right)=\frac{k_{1}r^{2}}{2}+W_{1},
\end{equation}
\noindent and 
\begin{equation}
\label{V2}
V_{2}\left( r\right)=\frac{k_{2}r^{2}}{2}-\frac{\alpha }{r^{2}}+W_{2}.
\end{equation}

The Schrodinger equation, in the Center-of-Mass (CM) system and in 
spherical coordinates, has the form (notation $\hbar=c=1$ is considered)
{\small \begin{equation}
\label{schr1}
\left\{ -\frac{1}{2\mu }\left\{ \frac{1}{r^{2}}\frac{\partial }
{\partial r}\left( r^{2}\frac{\partial }{\partial r}\right) +\frac{1}
{r^{2}}\left[ \frac{1}{\sin \theta }\frac{\partial }{\partial \theta }
\left( \sin \theta \frac{\partial }{\partial \theta }\right) +\frac{1}
{\sin ^{2}\theta }\frac{\partial ^{2}}{\partial \phi ^{2}}\right] 
\right\} 
-\left( E_{i}-V_{i}\right) \right\} \Psi _{i}\left( r,\theta ,\phi 
\right)=0,
\end{equation}}
\noindent where $i=1,2$ and $\mu =\frac{m_{1}m_{2}}{m_{1}+m_{2}}$ is the 
reduced mass.
Introducing the spherical harmonics
\begin{equation}
\label{wavfun1}
\Psi _{i,nl_{i}m_{i}}\left( r,\theta ,\phi \right) 
=\frac{1}{r}R_{i,nl_{i}}
\left( r\right) Y_{i,l_{i}m_{i}}\left( \theta ,\phi \right),
\end{equation}
\noindent the radial Schrodinger equation can be written as
\begin{equation}
\label{schr2}
\left\{ -\frac{1}{2\mu }\frac{d^{2}}{dr^{2}}+\frac{k_{i}r^{2}}{2}+
\frac{l_{i}\left( l_{i}+1\right) }{2\mu r^{2}}-\left( E_{i,nl_{i}}
-W_{i}\right) \right\} R_{i,nl_{i}}\left( r\right) =0,
\end{equation}
\noindent where
\[
l_{i}\left( l_{i}+1\right) =\left\{ \begin{array}{c}
l\left( l+1\right) \ \ \text{for} \ \ i=1, \\
l\left( l+1\right) -2\mu \alpha \ \ 
\text{for} \ \ i=2.
\end{array}\right. 
\]

The solution of Eq. (\ref{schr2}) can be found in any classical textbook of 
Quantum Mechanics \cite{Landau,Davydov} and has the form
\begin{equation}
\label{rad1}
R_{i,nl_{i}}\left( \xi \right) =N_{i,nl_{i}}e^{\frac{-\xi ^{2}}{2}}
\xi ^{l_{i}+1}F\left( -n,l_{i}+\frac{3}{2},\xi ^{2}\right),
\end{equation}
\noindent where
\[
\xi =\frac{r}{\sqrt{\frac{1}{\sqrt{\mu k}}}},
\]
\noindent $ N_{i,nl_{i}}$ is a normalization factor and $F\left( 
-n,l_{i}+\frac{3}{2},
\xi ^{2}\right) $ is the confluent hyper-geometric function. The 
eigenvalues of the energy are 
\begin{equation}
\label{ener1}
E_{i}=\sqrt{\frac{k}{\mu }}\left( 2n+l_{i}+\frac{3}{2}\right) +W_{i}.
\end{equation}

First, in order to compare with experimental values we fitted the 
parameters without imposing the flavor independence condition on the 
potentials. For that case \cite{Mart1} it is only possible to determine from 
the experimental spectra the values of $\varepsilon = \sqrt{\frac{k}{\mu 
}}$, $ 2\mu \alpha$ and $V_{0}=W_{0}+m_{1}+m_{2}$ that are shown in 
Table \ref{tab1} for charmonium ($c\overline{c}$) and upsilon 
($b\overline{b}$) systems. As it can be seen, the parameter $ 2\mu \alpha $ 
has the limiting value for the existence of the solution for the flavor 
dependent (FD) potential $ V_{2}$ for both quark systems, because
\[
l_{2}=-\frac{1}{2}+\sqrt{\left( l+\frac{1}{2}\right) ^{2}-2\mu \alpha},
\]
\noindent and for $l=0$ and $2\mu \alpha >0.25$ the squared root will 
have imaginary values.
\begin{table}[h]
\caption{Parameters obtained for the $c\overline{c}$ and
$b\overline{b}$ systems with the flavor dependent potential.}
\label{tab1}
\begin{tabular}{||c|c|c|c||}
\hline \hline 
System & $\varepsilon \left(MeV\right)$ 
&$2\mu \alpha$ &
$ V_{0}=W_{0}+m_{1}+m_{2}\left(MeV\right)$ \\ \hline \hline 
$V^{FD}_{1}$\footnote{Flavor dependent potential 
$V_{1}\left( r\right)$} ($c\overline{c}$)&296 & &2679 \\
$V^{FD}_{2}$\footnote{Flavor dependent potential 
$V_{2}\left( r\right)$} ($c\overline{c}$)&302 &0.25 &2773 \\
$V^{FD}_{1}$ ($b\overline{b}$)&217 & &9235 \\
$V^{FD}_{2}$ ($b\overline{b}$)&219 &0.25 &9301 \\ \hline \hline
\end{tabular}
\end{table}

\begin{figure}[h]
\begin{center}
\includegraphics[scale=0.47,angle=0]{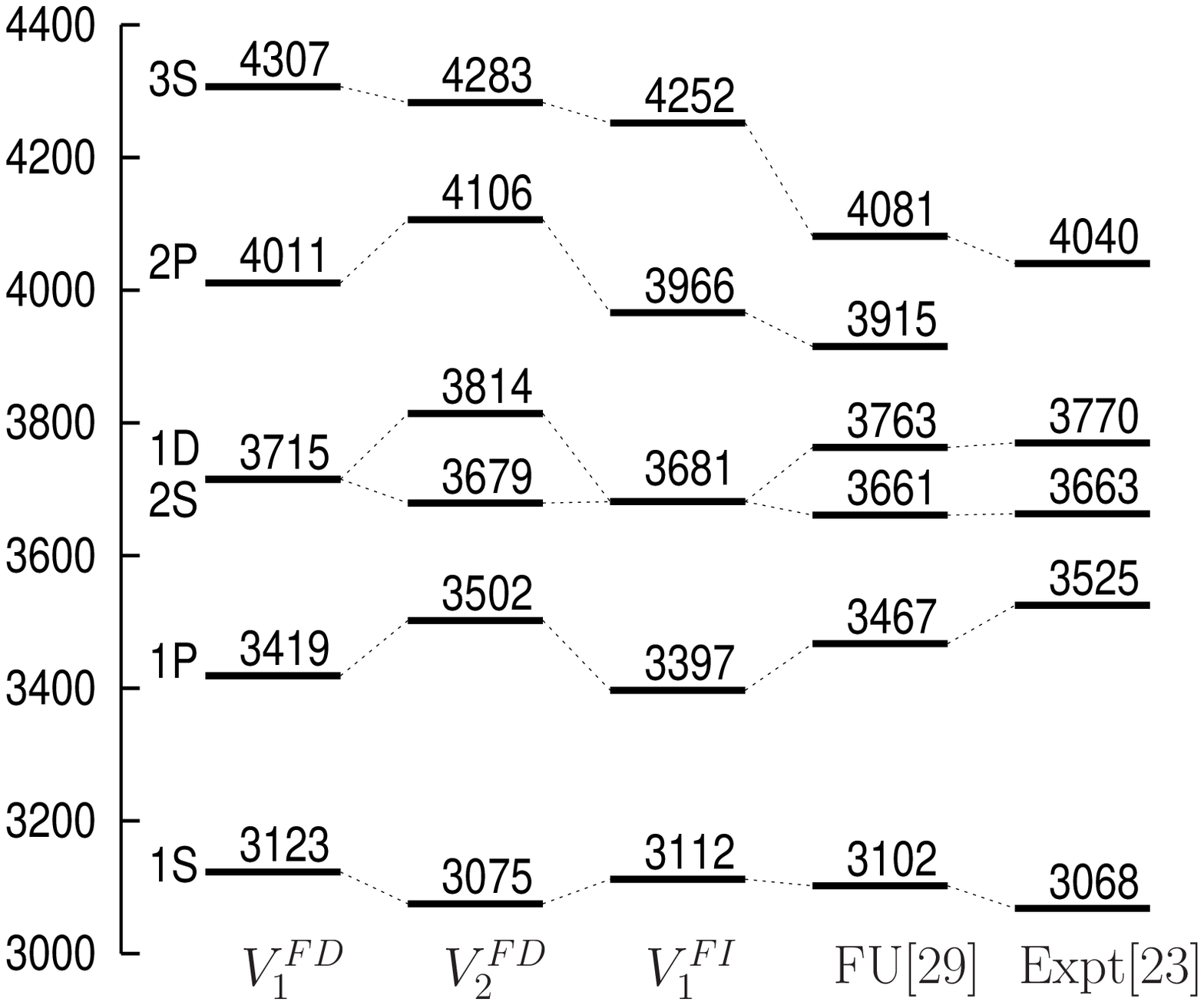}
\includegraphics[scale=0.47,angle=0]{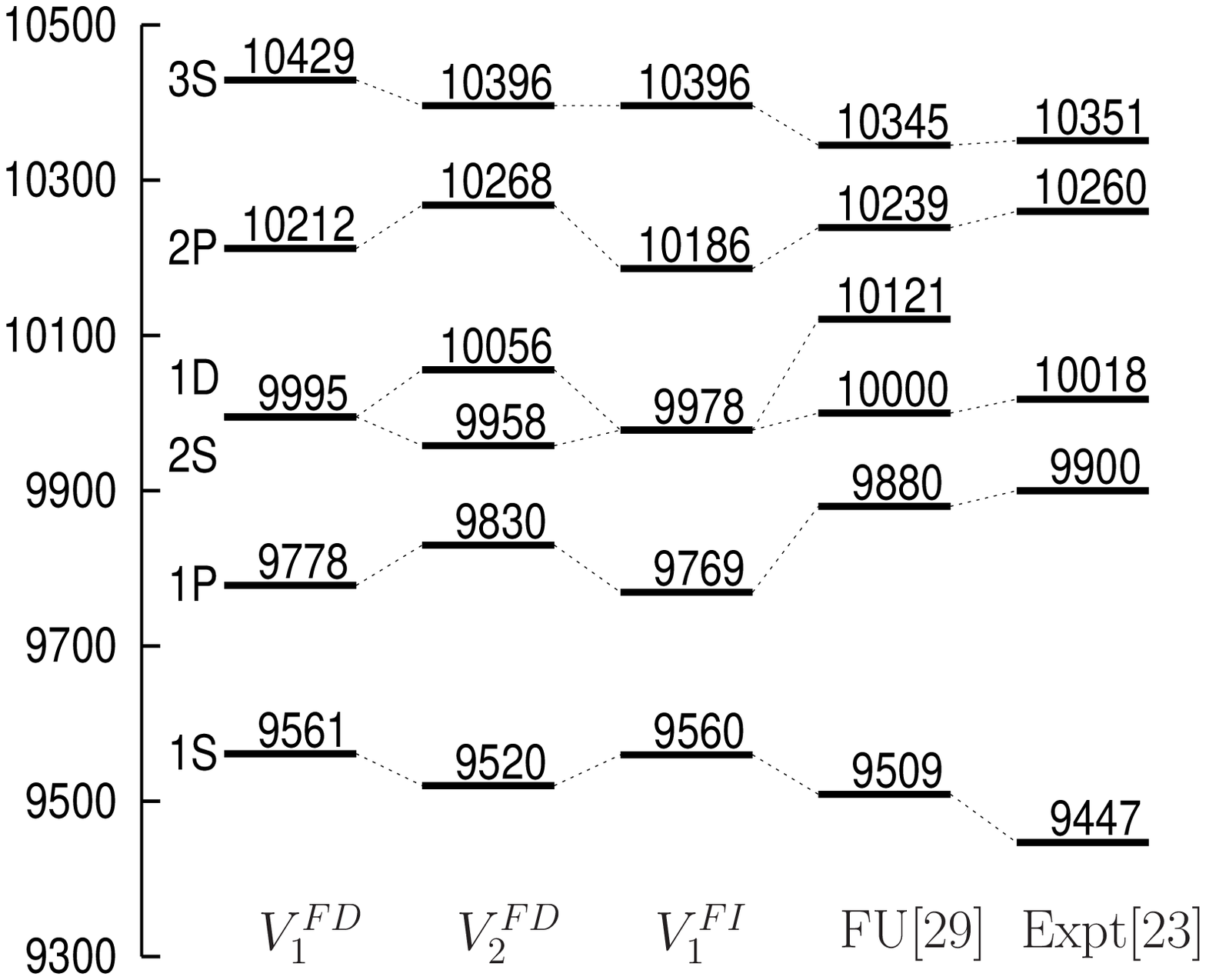}
\end{center}
\vspace{-0.7cm}
\caption{Charmonium (left) and Upsilom (right) energies ($MeV$) 
($V^{FI}_{1}$ is the flavor independent potential $V_{1}\left(r\right)$).}
\label{fig1}
\end{figure}

The mass values obtained for $c\overline{c}$ and $ b\overline{b}$ 
systems, with the parameters in Table \ref{tab1}, are presented in Figure
\ref{fig1} together with the theoretical values obtained by Fulcher 
\cite{Fulch3} and compared with the experimental values 
\cite{EuropR} (the results for each potential are presented in the figure
as columns). The experimental masses of the states $1S$, $1P$, $2S$ for 
the charmonium systems and all the ones presented for the Upsilom systems are 
spin averaged and were used in our fit of the parameters. The experimental 
masses reported for the $1D$ and $3S$ charmonium resonances belong to the 
$1^{3}\!D_{1}$ and $3^{3}\!S_{1}$ states. The results obtained for the potential 
$V_{2}$ are in better agreement with the experimental results because the 
non-harmonic term ($\sim \frac{1}{r^{2}}$) improves the behavior of the potential 
in the region $r\rightarrow 0$, and breaks some degeneracy present in the 
solution of the pure harmonic oscillator. Unfortunately the potential 
$V_{2}$ was found incompatible with the flavor independence condition, been 
impossible to obtain a unique potential like this for charmonium and 
upsilon systems. Then, after requiring the flavor independence condition only 
the pure harmonic oscillator parameters can be fitted, and the following 
values were obtained
\[
\begin{array}{cc}
k=0.155\ GeV^{3}, & m_{s}=2.725\ GeV,\\
W_{0}=-4.94\ GeV, & m_{c}=3.812\ GeV,\\ & m_{b}=7.093\ GeV,
\end{array}
\]
\noindent where the $s$ quark was included as in Martin's works. The 
spectra calculated with these parameters, for $c\overline{c}$ and 
$b\overline{b}$ mesons are presented in Figure \ref{fig1} ($V^{FI}_{1}$). 
Other meson resonances like  $s\overline{s}$, $c\overline{s}-s\overline{c}$, 
$b\overline{s}$ and $b\overline{c}$ are shown in Table \ref{tab4}. 
The mass of the $1S$ resonance for the $s\overline{s}$ meson 
was employed to fit the $s$ quark mass.

\begin{table}[h]
\caption{Other meson resonances ($MeV$).}
\label{tab4}
\begin{tabular}{||c|c|c|c|c|c|c|c|c|c||}
\hline \hline State&
$V^{FI}_{1}$($s\overline{s}$)& Expt\footnotemark[1]($s\overline{s}$)&
$V^{FI}_{1}$($c\overline{s}$)& Expt\footnotemark[1]($c\overline{s}$)&
$V^{FI}_{1}$($s\overline{b}$)& Expt\footnotemark[1]($s\overline{b}$)&
$V^{FI}_{1}$($b\overline{c}$)& Expt\footnotemark[1]($b\overline{c}$)&
FU\footnotemark[2]($b\overline{c}$)\\
\hline \hline 
1S&1016 & 1019&2065 & 2076& 5299& 5370&6340 & 6400& 
6361 \\
1P&1353 & &2377 & &5580 & &6590 & & 6703 \\
2S&1690 & &2689 & &5861 & &6840 & & 6876 \\ \hline \hline 
\end{tabular}
\footnotetext[1]{In Ref. \cite{EuropR1}.}
\footnotetext[2]{In Ref. \cite{Fulch3}.}
\end{table}

As it can be seen the theoretical results obtained are only estimates 
for the experimental values. The main differences between these 
theoretical results and the experimental ones, or other theoretical 
calculations, are clearly due to the non-singularity of the potential at the 
origin, and its concavity. That is, the harmonic potential has positive 
second derivative at variance with other proposed potentials that vary more 
slowly with the distance. This causes that when the energy of a state 
increases, the classical allowed region will be smaller for the harmonic 
potential than for the other ones and states result more localized. 
Other undesirable effects are the constant spacing between the consecutive 
levels and the degeneracy present. The differences between the proposed 
potential and the former ones also cause the so called constituent 
quarks masses to be here like twice of the usual values for the heavy 
quarks $c$, $b$ and fifth times for the $s$ quarks. The addition of the non 
harmonic term improved the results for the lower levels and broke the 
degeneracy present, but for the higher excited states the deficiencies 
remain; and this potential was also unable to fit with the flavor 
independence condition.

\section{Properties of the radial wave-function at the origin and 
related magnitudes \label{3}}

\begin{figure}[h]
\begin{center}
\includegraphics[scale=0.55,angle=0]{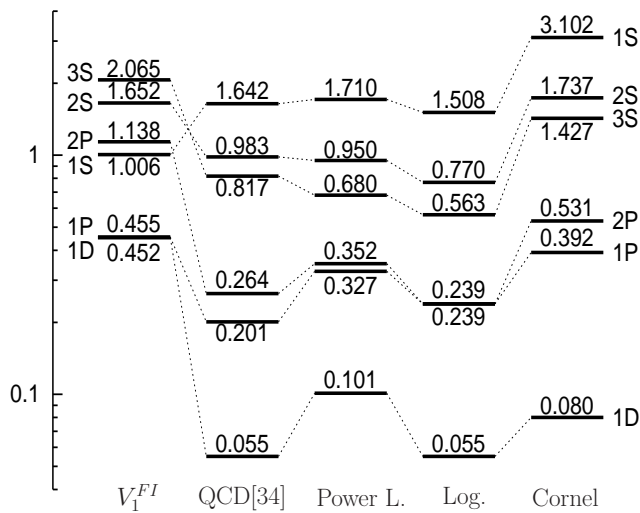}
\end{center}
\vspace{-0.7cm}
\caption{Radial wave functions at the origin and related quantities 
($\left| R_{nl}^{\left( l\right) }(0)\right|^{2}\left( GeV\right)^{2l+3}$) 
for $c\overline{b}$ mesons.}
\label{fig2}
\end{figure}

As it was mentioned in the previous section, the major reason for 
differences between our calculation for the pure harmonic oscillator and 
previous ones are the non-singularity of the potential at the origin, and 
its concavity. The value of the radial wave function or its first 
non-vanishing derivative at the origin
\begin{equation}
\label{rad2}
R_{nl}^{\left( l\right) }(0)\equiv \left. \frac{d^{l}R_{nl}\left( 
r\right)}
{dr^{l}}\right| _{r=0},
\end{equation}
\noindent is needed for the evaluation of pseudo-scalar decay constants 
and production rates through heavy-quark fragmentation 
\cite{Braat,Cheung}. In Figure \ref{fig2} we compare our results for $\left| 
R_{nl}^{\left( l\right) } (0)\right|^{2}$ with the ones presented for other 
potentials in Ref. \cite{Eich4} for $c\overline{b}$ mesons.

Two interesting magnitudes to evaluate, related with the wave functions 
at the origin, are the leptonic widths and the hyperfine splitting.  

\begin{figure}[h]
\begin{center}
\includegraphics[scale=0.55,angle=0]{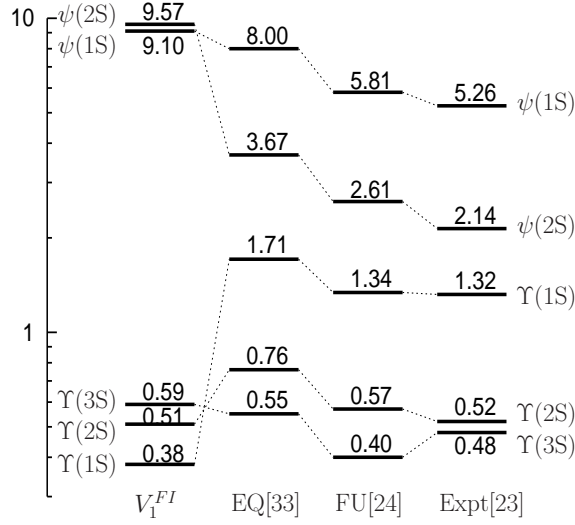}
\end{center}
\vspace{-0.7cm}
\caption{Leptonic widths ($KeV$).}
\label{fig3}
\end{figure}

Leptonic widths for charmonium and upsilon systems are presented in 
Figure \ref{fig3} and compared with results in Ref. \cite{Eich4,Fulch2}. 
They were obtained by the formula \cite{Royen}
\begin{equation}
\label{lept}
\Gamma \left( V^{0}\rightarrow e^{+}e^{-}\right) =\frac{16\pi 
N_{c}\alpha^{2}
e^{2}_{q}}{3}\frac{\left| \Psi \left( 0\right) \right| 
^{2}}{M^{2}_{V}},
\end{equation}
\noindent where $N_{c}=3$ (number of colors), $\alpha$ denotes the fine 
structure constant, $e_{q}$ denotes the quark charge and $M_{V}$ is the 
mass of the vector meson. 

\begin{figure}[h]
\begin{center}
\includegraphics[scale=0.55,angle=0]{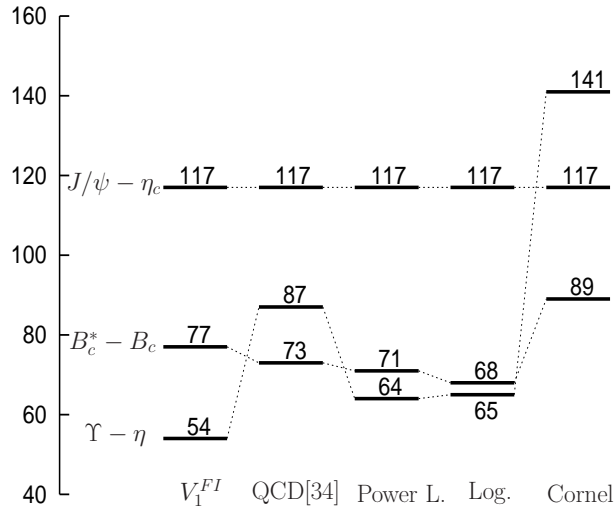}
\end{center}
\vspace{-0.7cm}
\caption{Hyperfine splitting for quarkonium ground states ($MeV$).}
\label{fig4}
\end{figure}

Finally, the hyperfine splitting can be obtained through the expression 
\cite{Mart1}

\begin{equation}
\label{hyp}
M\left( ^{3}S_{1}\right) -M\left( ^{1}S_{0}\right) =Cte\frac{\left| 
\Psi \left( 0\right) \right| ^{2}}{m_{a}m_{b}},
\end{equation}
\noindent in which the constant is fixed through the hyperfine 
splitting observed in the charmonium family
\[
M\left( J/\psi \right) -M\left( \eta _{c}\right) =117MeV.
\]
\noindent Results for the $c\overline{c}\  \left(J/\psi= ^{3}S_{1}
\ \eta_{c}=^{1}S_{0} \right)$, $b\overline{c}\  \left(B^{*}_{c}= ^{3}S_{1}
\ B_{c}=^{1}S_{0} \right)$ and $b\overline{b}\  \left(\Upsilon= ^{3}S_{1}
\ \eta=^{1}S_{0} \right)$
resonances are shown in Figure \ref{fig4}.

\section{Baryon Spectra\label{4}}

In the present section we study the baryon spectra using the harmonic 
potential obtained in Section \ref{2}. For the three-body system the 
Hamiltonian has the form 
\begin{equation}
\label{H1}
H=-\frac{1}{2m_{1}}\nabla _{r_{1}}^{2}-\frac{1}{2m_{2}}
\nabla _{r_{2}}^{2}-\frac{1}{2m_{3}}\nabla _{r_{3}}^{2}+V_{12}
\left( r_{12}\right) +V_{23}\left( r_{23}\right) +V_{31}\left( 
r_{31}\right). 
\end{equation}

In order to separate the C-M motion, we define the Jacobi 
($\overrightarrow{r}$, $\overrightarrow{R}$) and C-M ($\overrightarrow{R}_{CM}$) 
coordinates through
\begin{eqnarray}
\overrightarrow{r} & = & \left[ \frac{\mu _{bc}}{\mu _{a,bc}}\right] 
^{\frac{1}{4}}\left( 
\overrightarrow{r_{b}}-\overrightarrow{r_{c}}\right),\nonumber \\
\overrightarrow{R} & = & \left[ \frac{\mu _{a,bc}}{\mu _{bc}}\right] 
^{\frac{1}{4}}\left( \overrightarrow{r_{a}}-\frac{m_{b}
\overrightarrow{r_{b}}+m_{c}\overrightarrow{r_{c}}}{m_{b}+m_{c}}\right),\nonumber 
\\
\overrightarrow{R}_{CM} & = & \frac{m_{a}\overrightarrow{r_{a}}+m_{b}
\overrightarrow{r_{b}}+m_{c}\overrightarrow{r_{c}}}{M},\label{Jacobi1} 
\end{eqnarray}
\noindent where
\begin{eqnarray}
\mu _{bc} & = & \frac{m_{b}m_{c}}{m_{b}+m_{c}},\nonumber \\
\mu _{a,bc} & = & \frac{m_{a}\left( m_{b}+m_{c}\right) }{M},\nonumber 
\\
M & = & m_{a}+m_{b}+m_{c}.\label{mass1} 
\end{eqnarray}
\noindent Considering equal masses $m_{a}=m_{b}=m_{c}\equiv m$, and 
Eqs. (\ref{Jacobi1}), (\ref{mass1}) the C-M motion is separated and the 
Hamiltonian for the relative motion takes the form
\begin{equation}
\label{H2}
H=-\frac{1}{2\mu }\left( \nabla _{r}^{2}+\nabla _{R}^{2}\right) 
+\frac{\sqrt{3}}{4}k\left( 
\overrightarrow{r}^{2}+\overrightarrow{R}^{2}\right) 
+\frac{3}{2}W_{0},
\end{equation}
\noindent with 
\[
\mu \equiv \left( \frac{m_{a}m_{b}m_{c}}{M}\right)^{\frac{1}{2}}=
\frac{m}{\sqrt{3}},
\]
\noindent where the rule adopted by Richard \cite{Rich} (\ref{Rich}), 
was considered.

At this point we could obtain the baryon spectra directly from Eq. 
(\ref{H2}) noticing that it is the sum of two independent harmonic 
oscillators. But then the energy eigenvalues will not be in terms of the 
natural quantum numbers of the system, and it will not be possible a check 
with experimental or other theoretical results. The same problem is faced
with the usual 3-dimensional harmonic oscillator, which could be 
solved in Cartesian coordinates as the sum of three independent 1-dimensional 
harmonic oscillators, but then no relation between the conserved angular 
momentum and the energy spectrum is obtained and the use of spherical 
coordinates is convenient. 

Then introducing the hyper-spherical coordinates \cite{Dario} 
(see Appendix \ref{A})
\begin{eqnarray}
r_{x} & = & \rho \cos \left( \chi \right) \sin \left( \theta 
_{r}\right) 
\cos \left( \varphi _{r}\right), \nonumber \\
r_{y} & = & \rho \cos \left( \chi \right) \sin \left( \theta 
_{r}\right) 
\sin \left( \varphi _{r}\right), \nonumber \\
r_{z} & = & \rho \cos \left( \chi \right) \cos \left( \theta 
_{r}\right), 
\nonumber \\
R_{x} & = & \rho \sin \left( \chi \right) \sin \left( \theta 
_{R}\right) 
\cos \left( \varphi _{R}\right), \nonumber \\
R_{y} & = & \rho \sin \left( \chi \right) \sin \left( \theta 
_{R}\right) 
\sin \left( \varphi _{R}\right), \nonumber \\
R_{z} & = & \rho \sin \left( \chi \right) \cos \left( \theta 
_{R}\right), 
\label{hyper1} 
\end{eqnarray}
\noindent the kinetic term stay in a diagonal form and the potential 
become only dependent of the hyper-radio, then the Hamiltonian (\ref{H2}) 
takes the form
{\small
\begin{eqnarray}
H & = & -\frac{1}{2\mu }\left[ \frac{1}{\rho ^{5}}\frac{\partial }
{\partial \rho }\left( \rho ^{5}\frac{\partial }{\partial \rho }\right)
+\frac{1}{\rho ^{2}}\left( \frac{1}{\sin ^{2}\left(2 \chi \right) }
\frac{\partial }{\partial \chi }\left( \sin ^{2}\left(2 \chi \right) 
\frac{\partial }
{\partial \chi }\right) +\frac{\widehat{J}^{2}\left( \theta _{r},
\varphi _{r}\right) }{\cos ^{2}\left( \chi \right) 
}+\frac{\widehat{L}^{2}
\left( \theta _{R},\varphi _{R}\right) }{\sin ^{2}\left( \chi \right) }
\right) \right] \nonumber \\
 &  & +\frac{\sqrt{3}}{4}k\left(\rho^{2}\right) 
+\frac{3}{2}W_{0},\label{H3} 
\end{eqnarray}}
\noindent In which $\widehat{J}$ is the angular momentum of the 
subsystem $bc$ and $\widehat{L}$ is the angular momentum of particle $a$ 
respect to the C-M of the two body subsystem $bc$.
\begin{eqnarray}
\widehat{J}^{2} & = & -\frac{1}{\sin \left( \theta _{r}\right) }
\frac{\partial }{\partial \theta _{r}}\left( \sin \theta _{r}
\frac{\partial }{\partial \theta _{r}}\right) -\frac{1}
{\sin ^{2}\left( \theta _{r}\right) }\frac{\partial ^{2}}{\partial 
\varphi _{r}},
\nonumber \\
\widehat{L}^{2} & = & -\frac{1}{\sin \left( \theta _{R}\right) }
\frac{\partial }{\partial \theta _{R}}\left( \sin \theta 
_{R}\frac{\partial }
{\partial \theta _{R}}\right) -\frac{1}{\sin ^{2}\left( \theta _{R}
\right) }\frac{\partial ^{2}}{\partial \varphi _{R}}.\label{mom} 
\end{eqnarray}

The Schrodinger equation in this case has also analytical solution (see 
Appendix \ref{A}), with eigenvectors
\begin{eqnarray}
\Psi_{N,\lambda,j,m_{j},l,l_{j}} \left(\xi,\chi ,\theta _{r},\varphi 
_{r},\theta _{R},
\varphi _{R}\right)  
& = & N_{N,\lambda,j,l}e^{-\frac{\xi ^{2}}{2}}\xi ^{\lambda 
}L_{N}^{\lambda +2}
\left( \xi ^{2}
\right) \cos ^{j+\frac{1}{2}}\left( \chi \right) \sin 
^{l+\frac{1}{2}}\left( 
\chi \right)\times \nonumber \\
 &  & \times P_{\frac{\lambda 
-j-l}{2}}^{l+\frac{1}{2},j+\frac{1}{2}}\left( \cos 
\left( 2\chi \right) \right) Y^{m_{l}}_{l}\left( \theta _{R},\varphi 
_{R}
\right) Y^{m_{j}}_{j}\left( \theta _{r},\varphi _{r}\right), 
\label{barEig} 
\end{eqnarray}
\noindent where
\begin{equation}
\label{lamb}
\lambda =2n+j+l.
\end{equation}

\begin{figure}[h]
\begin{center}
\includegraphics[scale=0.49,angle=0]{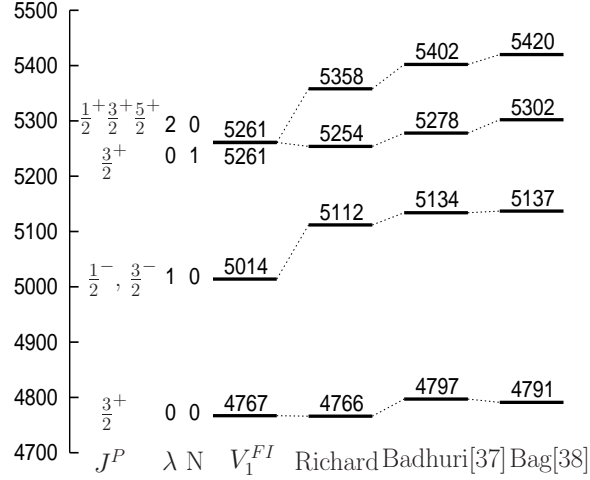}
\end{center}
\vspace{-0.7cm}
\caption{Baryon $ccc$ energies ($MeV$).}
\label{fig5}
\end{figure}

\begin{figure}[h]
\begin{center}
\includegraphics[scale=0.49,angle=0]{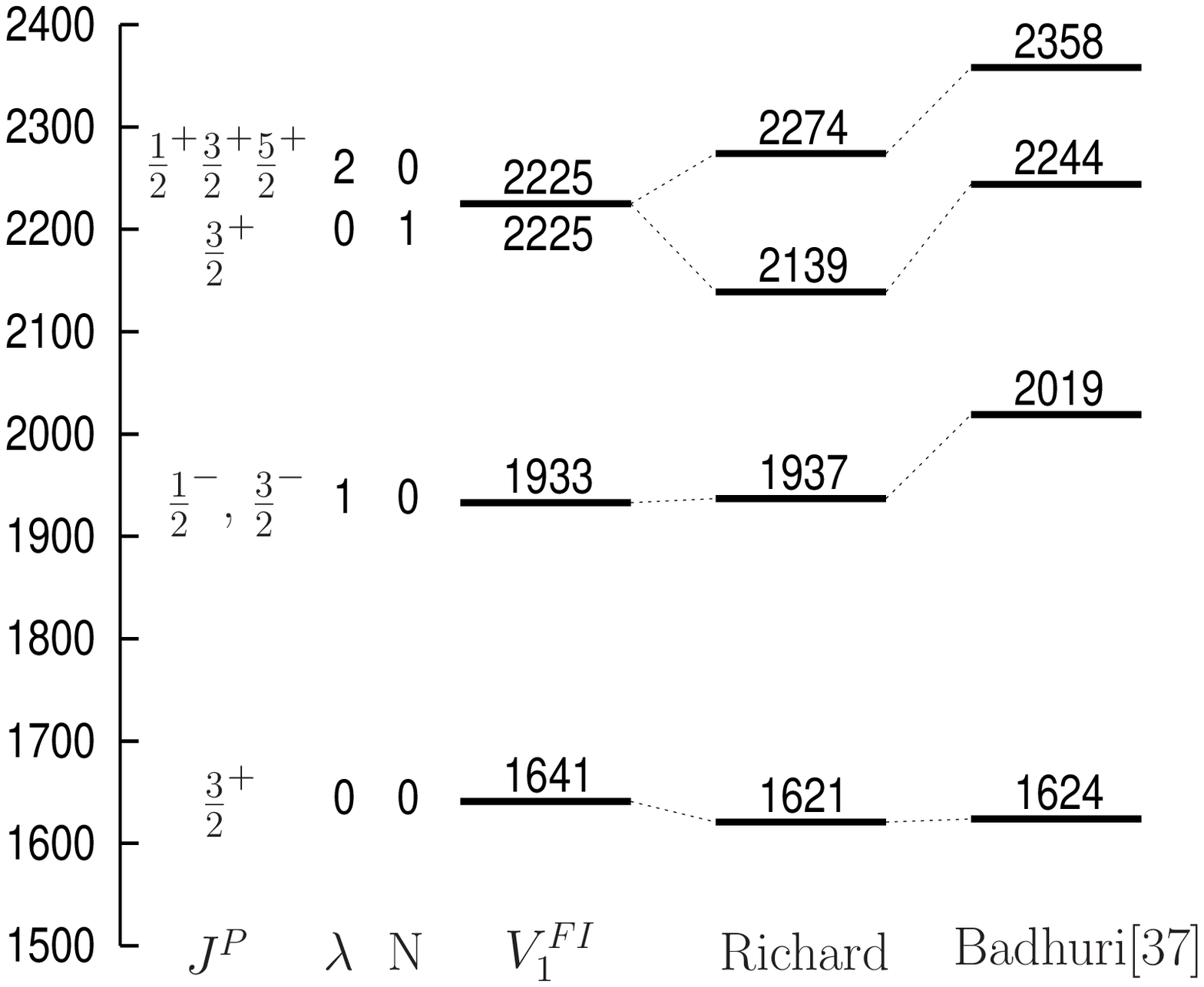}
\includegraphics[scale=0.49,angle=0]{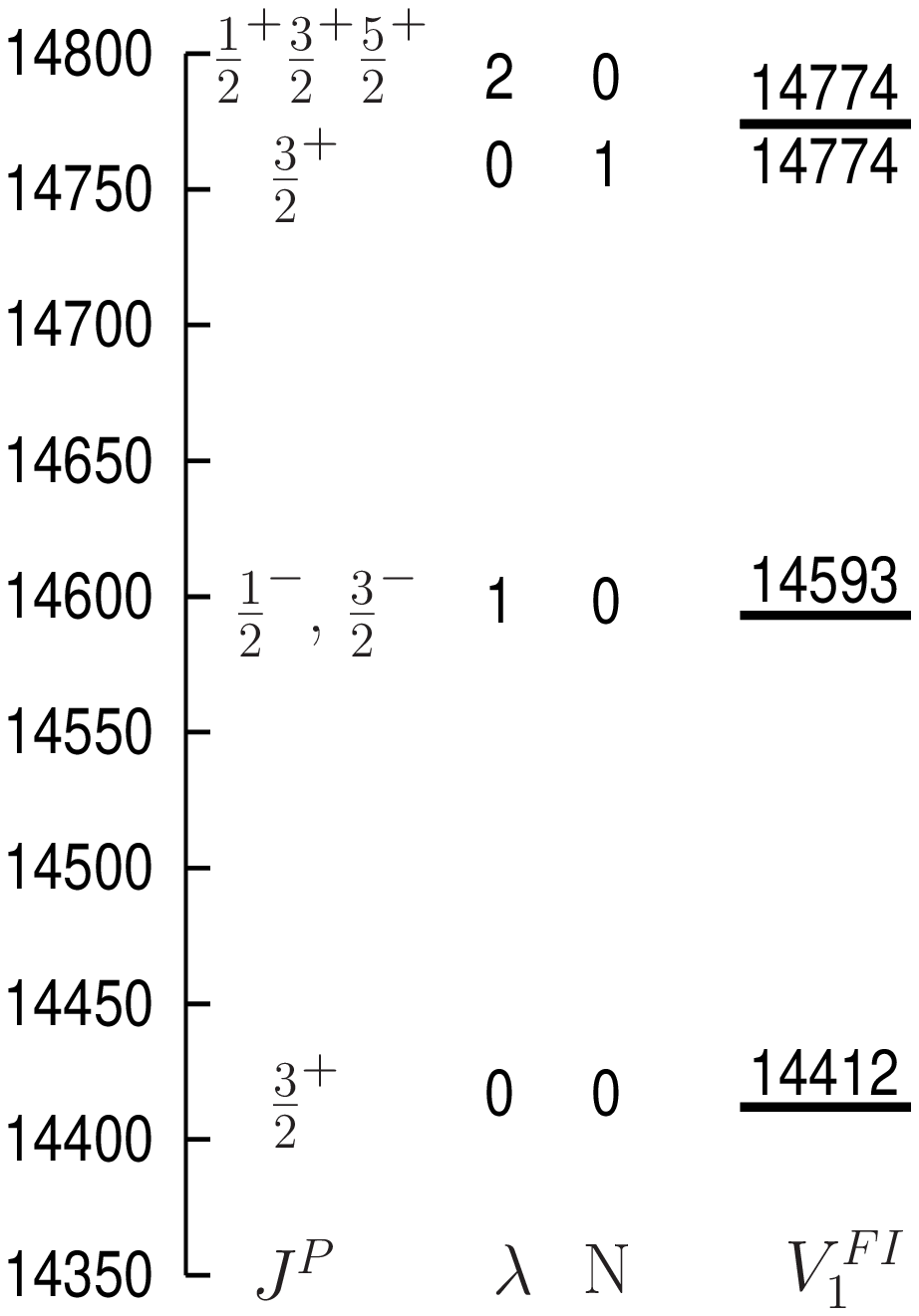}
\end{center}
\vspace{-0.7cm}
\caption{Baryon $sss$ (left) and $bbb$ (right) energies ($MeV$).}
\label{fig6}
\end{figure}

The eigenvalues are given by the expression
\begin{equation}
\label{barenr}
E=\sqrt{\frac{\sqrt{3}}{2}}\sqrt{\frac{k}{\mu}}\left( 2N+\lambda 
+3\right) 
+\frac{3}{2}W_{0},
\end{equation}
where $N$ is the number of nodes of the hyper-radial function, 
$\lambda$ is the grand-angular quantum number, and $P_{\frac{\lambda -j-l} 
{2}}^{l+\frac{1}{2},j+\frac{1}{2}}\left( \cos \left( 2\chi \right) \right)$
are the Jacobi polynomials. In Figures  \ref{fig5} and \ref{fig6} we compare 
our results with other calculations presented by Richard in Ref \cite{Rich}.

\vspace{-0.7cm}

To compare with experiments there were only two equal constituent quark 
baryons at hand \cite{EuropR1}, for the $sss$ system, for which
\[
\begin{array}{cc}
m\left( \Omega ^{-}\right) _{Expt}=1672\ MeV, & m\left( \Omega 
^{-}\right) 
_{Expt}=2250 MeV,\\
m\left( \Omega ^{-}\right) _{V^{FI}_{1}}=1641\ MeV, & m\left( \Omega 
^{-}\right) 
_{V^{FI}_{1}}=2225\ MeV, \\
\frac{m\left( \Omega ^{-}\right) _{Expt}-m\left( \Omega ^{-}\right) 
_{V^{FI}_{1}}}{m\left( \Omega ^{-}\right) _{Expt}}=0.02, & 
\frac{m\left( \Omega ^{-}\right) _{Expt}-m\left( \Omega ^{-}\right) 
_{V^{FI}_{1}}}{m\left( \Omega ^{-}\right) _{Expt}}=0.01.
\end{array}
\]

\section{Conclusions}

In the present paper we have studied, within a non-relativistic 
spin-independent model with harmonic confining potential, the spectra and 
other properties of hadron systems.

It was found that for mesons, without imposing the flavor independence 
condition, two possible potentials with harmonic confining terms had 
analytical solutions that give good estimates of the experimental values 
reported for the meson spectra. The better fit was obtained for the 
potential with a term proportional to $\frac{1}{r^{2}}$ because it has a 
singularity for $r\rightarrow 0$ that improves its behavior in this 
region. However this potential was found to be incompatible with the flavor 
independence condition and was not considered in the analysis that 
followed.

For the pure harmonic oscillator the parameters introduced were fixed 
from the low lying levels of heavy quarks systems and imposing the 
flavor independence condition. The calculation of the meson and baryon 
spectra, and the hyperfine splitting with this potential give 
good estimates of the experimental and other theoretical results; in 
the case of leptonic widths we could say that the results are not good.
Although this potential is far from being a good approximation for the 
real inter-quark potential, and the results are not as good as the 
theoretically obtained by other phenomenological models, it has the great 
advantage that allows to obtain analytical solutions for both meson and 
baryon spectra. That is, reasonable theoretical results are obtained 
without the need of numerical methods and computational calculations. The 
major differences of this potential and the others mentioned in the 
introduction are due to its non-singularity at the origin and its 
concavity, that cause the bad results for the obtained leptonic widths and 
also (for a better fit with experiments) the so called 
constituent masses to be bigger than the usual ones. The baryon spectra 
was studied with the use of the rule (\ref{Rich}) for the $qq$ potential 
and for obtaining the analytical solution it was necessary to restrict 
the study to equal constituent quark systems.

We finalize with the conclusions of A. Martin in Ref. \cite{Mart3} 
``... if you want to know the mass of a particle and if you have a little 
time (in years!) and little money you should forget all your prejudices 
and use potential models''.

\appendix
\section{The Hyper-spherical coordinates and the solution of the 
three-body problem with a harmonic potential \label{A}}

The hyper-spherical coordinates are very useful for dealing with the 
three-body problem; in what follows we make a small review of them. 

The kinetic energy of the Hamiltonian (\ref{H2}),
\begin{equation}
\label{a1}
\widehat{K}=-\frac{1}{2\mu }\left( \nabla _{r}^{2}+\nabla 
_{R}^{2}\right) ,
\end{equation}
\noindent can be written as a Laplacian in a 6-dimensional space, due 
to the symmetry in the two Jacobi vectors. When a change of coordinates 
is made, the $N$-dimensional Laplacian transforms as
\begin{equation}
\label{a2}
\Delta =\frac{1}{\prod\limits_{i}l_{i}}\sum ^{N}_{k=1}\frac{\partial 
}{\partial x_{k}'}
\left( \frac{\prod\limits_{j}l_{j}}{l^{2}_{k}}\frac{\partial }{\partial 
x_{k}'}\right) 
\ \ \text{where}\ \ l_{j}=\sqrt{\sum ^{N}_{i=1}\left( \frac{\partial 
x_{i}}
{\partial x_{j}'}\right)^{2}} ,
\end{equation}
\noindent are the metric coefficients. The change to hyper-spherical 
coordinates is based on the definition of the hyper-radius by
\begin{equation}
\label{a3}
\rho =\sqrt{\sum ^{N}_{i=1}x^{2}_{i}},
\end{equation}
\noindent and $N-1$ angles in a way that (\ref{a3}) is satisfied and 
the old variables are expressed in terms of the new ones by $N$ functions 
with the form
\begin{equation}
\label{a4}
x_{i}=\rho F_{i}\left( \Omega _{N-1}\right) ,
\end{equation}
\noindent and the Laplacian operator becomes
\begin{equation}
\label{a5}
\Delta =\frac{1}{\rho ^{N-1}}\frac{\partial }{\partial \rho }\left( 
\rho ^{N-1}
\frac{\partial }{\partial \rho }\right) +\frac{1}{\rho ^{2}}\Delta 
_{\Omega _{N-1}}.
\end{equation}

In the expression (\ref{a5}) $\Omega _{N-1}$ denotes all angles. The 
explicit form of the angular term of the Laplacian operator and its 
eigenfunctions will depend of the set of angles selected as new coordinates 
and the eigenvalues will be equal to $-\lambda \left( \lambda 
+N-2\right) $. In this 6-dimensional case where the selected coordinates are 
(\ref{hyper1}) the angular term obtained is
\begin{equation}
\label{a6}
\Delta _{\Omega _{N-1}}=\frac{1}{\sin ^{2}\left( 2\chi \right) 
}\frac{\partial }
{\partial \chi }\left(\sin ^{2}\left(2\chi \right)\frac{\partial 
}{\partial \chi }\right)+
\frac{1}{\cos ^{2}\left( \chi \right) }\Delta _{\theta _{r},\varphi 
_{r}}+\frac{1}
{\sin ^{2}\left( \chi \right) }\Delta _{\theta _{R},\varphi _{R}}.
\end{equation}

Its eigenfunctions are expressed as a product of orthogonal polynomials 
in separated variables \cite{Dario}. The eigenfunctions corresponding 
to Jacobi's angles $\theta _{r}$, $\varphi _{r}$, $\theta _{R}$ and 
$\varphi _{R}$ are the well known spherical harmonics. Then the equation 
obtained for the function of the angle $\chi$ is
\begin{equation}
\label{a7}
\left[ \frac{1}{\sin ^{2}\left( 2\chi \right) }\frac{\partial 
}{\partial \chi }\left( 
\sin ^{2}2\chi \frac{\partial }{\partial \chi }\right) -\frac{l\left( 
l+1\right) }{
\sin^{2}\left( \chi \right) }-\frac{j\left( j+1\right) }{\cos 
^{2}\left( \chi \right) 
}\right] F^{l,j}_{\lambda }\left( \chi \right) =-\lambda \left( \lambda 
+4\right) 
F^{l,j}_{\lambda }\left( \chi \right), 
\end{equation}
\noindent for which the solution are the Jacobi polynomials
\begin{equation}
\label{a8}
F^{l,j}_{\lambda }\left( \chi \right) =N_{\lambda ,l,j}\sin 
^{l+\frac{1}{2}}\left( 
\chi \right) \cos ^{j+\frac{1}{2}}\left( \chi \right) P_{\frac{\lambda 
-j-l}{2}}^{l+
\frac{1}{2},j+\frac{1}{2}}\left( \cos \left( 2\chi \right) \right), 
\end{equation}
\noindent with $\lambda =2n+l+j$ and $N_{\lambda ,l,j}$ a normalization 
factor.

For the harmonic interaction between equal mass particles, the 
potential is only dependent of the hyper-radius, then we can separate variables 
and the radial equation has the form
\begin{equation}
\label{a9}
\left[ -\frac{1}{2\mu \rho ^{5}}\frac{\partial }{\partial \rho }\left( 
\rho ^{5}
\frac{\partial }{\partial \rho }\right) +\frac{\lambda \left( \lambda 
+4\right) 
}{2\mu \rho ^{2}}+\frac{\sqrt{3}k}{4}\rho ^{2}-\left( E_{N,\lambda 
}-\frac{3}{2}W_{0}\right) 
\right] R_{N,\lambda }\left( \rho \right) =0.
\end{equation}

Introducing new variables
\begin{equation}
\label{a10}
\xi =\left( \frac{\rho }{\rho _{0}}\right) \sqrt{\beta },\ \ \ \rho 
_{0}=\left( \frac
{1} {2\mu \left( E-\frac{3}{2}W_{0}\right) }\right) ^{\frac{1}{2}},\ \ 
\ \beta ^{2}=
\frac{\left( k\sqrt{3}\right) }{8\mu \left( E-\frac{3}{2}W_{0}\right) 
^{2}},
\end{equation}
\noindent and the new function
\begin{equation}
\label{a11}
T_{N,\lambda }=\frac{R_{N,\lambda }}{\xi ^{\frac{5}{2}}},
\end{equation}
\noindent we obtain for (\ref{a9}) the equation
\begin{equation}
\label{a12}
\left[ \frac{d^{2}}{d\xi ^{2}}+\frac{1}{\beta _{N,\lambda 
}}-\frac{\lambda \left( 
\lambda +4\right) +\frac{15}{4}}{\xi ^{2}}-\xi ^{2}\right] T_{N,\lambda 
}\left( \xi \right) =0,
\end{equation}
\noindent with solutions
\begin{equation}
\label{a13}
T_{N,\lambda }\left( \xi \right) =e^{-\frac{\xi ^{2}}{2}}\xi ^{\lambda 
+\frac{5}{2}}L_{N}^
{\lambda +2}\left( \xi ^{2}\right), 
\end{equation}
\noindent where 
\begin{equation}
\label{a14}
\frac{1}{\beta _{N,\lambda }}=4s+2\left( \lambda +2\right) +2,
\end{equation}
\noindent and $L_{N}^{\lambda +2}\left( \xi ^{2}\right)$ are the 
generalized Laguerre polynomials.

Then the complete eigenvectors and the eigenvalues of the problem are 
given by Eqs. (\ref{barEig}) and (\ref{barenr}).

\begin{acknowledgments}
One of us (M.R.) would like to thank A. Dimarco, S. Duarte, and the 
Department of Nuclear and High Energy Physics at ``Centro Brasileiro de 
Pesquisas F\'{\i}sicas (CBPF)'' for their kind hospitality and useful 
discussions.  
\end{acknowledgments}


\end{document}